\documentclass[aps,preprint,showpacs,showkeys]{revtex4}
\usepackage{amsmath}
\usepackage{bm}

\input{tcilatex}
\begin{document}

\title{Solution of the Bosonic and Algebraic Hamiltonians by using AIM}
\author{Ramazan Ko\c{c}, Hayriye T\"{u}t\"{u}nc\"{u}ler and Eser Ol\u{g}ar}
\affiliation{Gaziantep University, Department of Physics, Faculty of Engineering 27310
Gaziantep/Turkey}
\email{koc@gantep.edu.tr, tutunculer@gantep.edu.tr, olgar@gantep.edu.tr.}
\date{\today }

\begin{abstract}
We apply the notion of asymptotic iteration method (AIM) to determine
eigenvalues of the bosonic Hamiltonians that include a wide class of quantum
optical models. We consider solutions of the Hamiltonians, which are even
polynomials of the fourth order with the respect to Boson operators. We also
demonstrate applicability of the method for obtaining eigenvalues of the
simple Lie algebraic structures. Eigenvalues of the multi-boson Hamiltonians
have been obtained by transforming in the form of the single boson
Hamiltonian in the framework of AIM.
\end{abstract}

\maketitle

\section{Introduction}

The study of the same problems from different point of view lead to the
progress of the science and include a lot of mathematical tastes. An
iteration technique \cite%
{ciftci1,sous,ciftci2,ciftci3,barakat1,barakat2,fernandez} has recently been
suggested to obtain eigenvalues of Schr\"{o}dinger equation which improves
both analytical and numerical determination of the eigenvalues and has been
developed for some matrix Hamiltonians, arising from the development of fast
computers\cite{koc1,koc2}. Asymptotic iteration method (AIM) is very
efficient to establish eigenvalues of the various quantum mechanical
systems, because of their simplicity and low round off error. This method
has been widely applied for determination of eigenvalues of the Schr\"{o}%
dinger type equations. Encouraged by its satisfactory performance through
comparisons with the other methods, we feel tempted to develop AIM to obtain
eigenvalues of algebraic Hamiltonians. In contrast to the solution of the
Schr\"{o}dinger equation by using AIM including Coulomb, Morse, harmonic
oscillator, etc. type potentials, the study of the algebraic Hamiltonians 
\cite{tutunculer,koc3,koc4} has not attracted much attention in the
literature. Such Hamiltonians have been found to be useful in the study of
electronic properties of semiconductors, quantum dots and quantum wells. It
is evident that the formalism can also be developed for solving algebraic
equations.

The algebraic techniques have been proven to be useful in the description of
the physical problems in a variety of fields \cite%
{koc3,koc4,qu,kara1,kara2,alv,klim,ocak}. In recent years there has been a
great deal of interest in quantum optical models which reveal new physical
phenomena described by the Hamiltonians expressed as nonlinear functions of
Lie algebra generators or boson and/or fermion operators\cite%
{koc5,koc6,koc7,rashba,tutunculer1}. Such systems have often been analyzed
by using numerical methods, because the implementation of the Lie algebraic
techniques to solve those problems is not very efficient and most of the
other analytical techniques do not yield simple analytical expressions. They
require tedious calculations. In principle, if a Hamiltonian is expressed by
boson operators, one could rely directly on the known formulae of the action
of boson operators on a state with a defined number of particles without
solving differential equations. Apart from the mentioned method, sometimes
the Hamiltonians can be put in a simple form by using the transformation
properties of the bosons.

In this article, AIM is suggested and adapted to solve the bosonic
Hamiltonians. We note that this has never been done before. As a particular
case our model includes the solutions of the Hamiltonian of the multiphoton
interactions and the Hamiltonian of the systems of photons and bosons
expressed in a single mode form. We briefly discuss the bosonic construction
of the various Hamiltonians. These Hamiltonians are not only mathematically
interesting but they also have potential interest in physics.

The paper is organized as follows. In section 2, we briefly review the
properties of boson and its differential realization. The procedure for
solving a bosonic Hamiltonian in the framework of the AIM is presented in
this section. Section 3 is devoted to illustrate determination of \ the
eigenvalues of a bosonic Hamiltonian in the framework of the AIM. The
bosonization of the physical Hamiltonians whose original forms are given as
differential operators is discussed. As a practical example we illustrate
the solution of the anharmonic oscillator and multiphoton interaction
problems. In section 4, we introduce a technique to obtain eigenvalues of
the two mode bosonic Hamiltonians by using AIM. We present the application
of the AIM in order to obtain eigenvalues for a class of models describing
two-mode multiphoton processes. Finally, we comment on the validity of our
method and remark on the possible use of our method in the different fields
of the physics.

\section{Basic Formalism and solution of single boson Hamiltonian}

In this section, we illustrate solution of the single boson Hamiltonians, by
modifying AIM. The usual differential realization of the annihilation
operator $a,$ and creation operator $a^{+},$ are given by 
\begin{equation}
a^{+}=\frac{1}{\sqrt{2}}\left( -\frac{\partial }{\partial x}+x\right) ;\ a=%
\frac{1}{\sqrt{2}}\left( \frac{\partial }{\partial x}+x\right)   \label{ad1}
\end{equation}%
and they act on the state $\left\vert n\right\rangle :$%
\begin{equation}
a^{+}\left\vert n\right\rangle =\sqrt{n+1}\left\vert n+1\right\rangle ;\
a\left\vert n\right\rangle =\sqrt{n}\left\vert n-1\right\rangle   \label{ad3}
\end{equation}%
with the commutation relation%
\begin{equation*}
\left[ a,a^{+}\right] =1.
\end{equation*}%
A single boson Hamiltonian describing a physical system can be expressed as 
\begin{equation}
H=\dsum\limits_{i}\gamma _{i,i}a^{i}(a^{+})^{i}+\dsum\limits_{i,j(i\neq
j)}\gamma _{i,j}a^{i}(a^{+})^{j}  \label{ad2}
\end{equation}%
where $\gamma _{i,j}$ is a constant. It is obvious that first part of then $H
$ is diagonal and exactly solvable. Second part of the Hamiltonian $H$
includes non-diagonal terms and it is usually solved by using various
perturbation techniques. Our task is now to develop an AIM to obtain
eigenvalues of $H.$ We assume that action of $H$ on the state $\left\vert
n\right\rangle $\ produce the following three term recurrence relation (or
reduced to three term recurrence relation) such that%
\begin{equation}
\left\vert n+2\right\rangle =r_{n}\left\vert n+1\right\rangle
+s_{n}\left\vert n\right\rangle   \label{ad4}
\end{equation}%
where $r_{n}$ and $s_{n}=E-s_{n}^{\prime }$ are function of $n$. From the
analogy of the AIM \cite{ciftci1} it follows that (\ref{ad4}) can be put a
more suitable form in order to obtain eigenvalues $E$ of $H$. Reformulation
of (\ref{ad4}) provides the following equations:%
\begin{eqnarray}
n &=&0;\ \left\vert 2\right\rangle =r_{0}\left\vert 1\right\rangle
+s_{0}\left\vert 0\right\rangle =p_{0}\left\vert 1\right\rangle
+q_{0}\left\vert 0\right\rangle   \notag \\
n &=&1;\ \left\vert 3\right\rangle =r_{1}\left\vert 2\right\rangle
+s_{1}\left\vert 1\right\rangle =p_{1}\left\vert 1\right\rangle
+q_{1}\left\vert 0\right\rangle   \notag \\
&&\cdots   \label{ad5} \\
n &=&m;\ \left\vert m+2\right\rangle =r_{m}\left\vert m+1\right\rangle
+s_{m}\left\vert m\right\rangle =p_{m}\left\vert 1\right\rangle
+q_{m}\left\vert 0\right\rangle   \notag
\end{eqnarray}%
where $p_{m}$ and $q_{m}$ are given by%
\begin{eqnarray}
p_{m} &=&r_{m}p_{m-1}+s_{m}p_{m-2}  \notag \\
q_{m} &=&r_{m}q_{m-1}+s_{m}q_{m-2}  \label{ad7}
\end{eqnarray}%
with the initial conditions%
\begin{equation*}
p_{-1}=q_{-2}=1\ \text{and\ }p_{-2}=q_{-1}=0.
\end{equation*}%
To this end we assume that $m$ is large enough and the states reach their
asymptotic values. Thus we can write%
\begin{eqnarray}
\left\vert m+2\right\rangle  &=&p_{m}\left\vert 1\right\rangle
+q_{m}\left\vert 0\right\rangle   \notag \\
\left\vert m+3\right\rangle  &=&p_{m+1}\left\vert 1\right\rangle
+q_{m+1}\left\vert 0\right\rangle   \label{ad71}
\end{eqnarray}%
After all we can concisely write that 
\begin{equation}
\frac{p_{m}}{q_{m}}=\frac{\ p_{m+1}}{q_{m+1}}\ \text{or\ }%
q_{m}p_{m+1}-q_{m+1}p_{m}=0.  \label{ad6}
\end{equation}%
The last equation can be solved for eigenvalues $E$, then the last
approximation leads to the determination of the eigenvalues of the
Hamiltonian $H$. Before going futher, we note that eigenvalues of the
associated problem can be obtained by using the following MATHEMATICA
program code. Let us define $\left\vert n\right\rangle =f[n]$ then

\QTP{Body Math}
k = 20; Do[f[n+2] = Simplify[r$_{n}$f[n+1]+s$_{n}$f[n]], \{n, 0, k\}]

\QTP{Body Math}
$\qquad $(*where k is number of iteration*)

\QTP{Body Math}
NSolve[Coefficient[f[k + 2], f[0]]*Coefficient[f[k ], f[2]] -

\QTP{Body Math}
$\qquad \ \ \ \ \ \ $Coefficient[f[k + 2], f[2]]*Coefficient[f[k ], f[0]] ==
0, E1]

\QTP{Body Math}
$\qquad $(*E1 is eigenvalues of the H*)

In the next sections, we want to illustrate our task on an explicit example.

\section{Eigenstate of the single boson Hamiltonians}

In this section we study the determination of the single and multi-boson
Hamiltonians in the framework of the AIM.

\subsubsection{Anharmonic oscillator}

The solution of the Schr\"{o}dinger equation including anharmonic potential
has attracted a lot of attention, arising its considerable impact on the
various branches of physics as well as biology and chemistry. Besides its
importance in physics, biology and chemistry, in practice anharmonic
oscillator problem is always used to test the accuracy and the efficiency of
the unperturbative methods. In this section we take a new look at the
solution of the anharmonic oscillator problem through the modified AIM. The
equation is described by the Hamiltonian: 
\begin{equation}
H=-\frac{d^{2}}{dx^{2}}+x^{2}+\alpha x^{4}.  \label{eq5}
\end{equation}%
where $\alpha $ is a constant. Our task is now to demonstrate that the
Hamiltonian (\ref{eq5}) can be expressed in terms of the bosons. One way to
express the Hamiltonian $H$ with boson operators is to use an appropriate
differential realization of bosons. Using the realization (\ref{ad1}), the
Hamiltonian (\ref{eq5}) can be written as: 
\begin{equation}
H=a^{+}a+aa^{+}+\frac{\alpha }{4}(a+a^{+})^{4}.  \label{eq:6}
\end{equation}%
When the Hamiltonian (\ref{eq:6}) acts on the state $\left\vert
n\right\rangle $, the eigenvalue equation $H\left\vert n\right\rangle
=E\left\vert n\right\rangle $ can be transformed to the following recurrence
relation:%
\begin{eqnarray}
( &&H-E)\left\vert n\right\rangle =\left( 2n+1-E\right) \left\vert
n\right\rangle +\frac{3\alpha }{2}\left( n+n^{2}+\frac{1}{2}\right)
\left\vert n\right\rangle +  \notag \\
&&\alpha \sqrt{(n+1)(n+2)}\left( n+\frac{3}{2}\right) \left\vert
n+2\right\rangle +\alpha \sqrt{n(n-1)}\left( n-\frac{1}{2}\right) \left\vert
n-2\right\rangle +  \notag \\
&&\frac{\alpha }{4}\sqrt{(n+1)(n+2)(n+3)(n+4)}\left\vert n+4\right\rangle +%
\frac{\alpha }{4}\sqrt{n(n-1)(n-2)(n-3)}\left\vert n-4\right\rangle =0.
\label{eq:7}
\end{eqnarray}%
Here, the skill is to express the $n^{th}$ \textit{even} state in terms of $%
\left\vert 0\right\rangle $ and $\left\vert 2\right\rangle $ states and $%
n^{th}$ \textit{odd} state in terms of $\left\vert 1\right\rangle $ and $%
\left\vert 3\right\rangle $ states. Applying the technique given in the
previous section, we can obtain the following expressions:%
\begin{eqnarray}
n &=&0;\ \left\vert 4\right\rangle =p_{0}\left\vert 0\right\rangle
+q_{0}\left\vert 2\right\rangle  \notag \\
n &=&2;\ \left\vert 6\right\rangle =p_{2}\left\vert 0\right\rangle
+q_{2}\left\vert 2\right\rangle  \notag \\
&&\cdots  \label{eq:8} \\
n &=&m;\ \left\vert m+4\right\rangle =p_{m}\left\vert 0\right\rangle
+q_{m}\left\vert 2\right\rangle  \notag \\
n &=&m+2;\ \left\vert m+6\right\rangle =p_{m+2}\left\vert 0\right\rangle
+q_{m+2}\left\vert 2\right\rangle .  \notag
\end{eqnarray}%
The truncation of the state for large values of $m$ leads to the following
relations%
\begin{equation}
q_{m}p_{m+2}-p_{m}q_{m+2}=0.  \label{eq:10}
\end{equation}%
Here $p_{i}$ and $q_{i}$ can be calculated by using the following
MATHEMATICA program code (again we define $\left\vert n\right\rangle =f[n]$)

\QTP{Body Math}
s1 = Collect[Simplify[Solve[(H-E)f[n] == 0, f[n + 4]]], \{f[n\_]\}]

\QTP{Body Math}
$\qquad $(*f[n+4] is obtained from (\ref{eq:7})*)

\QTP{Body Math}
k = 20; Do[f[n+4] = Simplify[s1[[1,1,2]]], \{n, 0, k\}]

\QTP{Body Math}
$\qquad $(*where k is number of iteration*)

\QTP{Body Math}
Solve[Coefficient[f[k + 4], f[0]]*Coefficient[f[k + 2], f[2]] -

\QTP{Body Math}
$\qquad \ \ \ $Coefficient[f[k + 4], f[2]]*Coefficient[f[k + 2], f[0]] == 0,
E1] /. $\alpha \rightarrow $ 0.1

\QTP{Body Math}
$\qquad $(*gives eigenvalues of the even state*)

\QTP{Body Math}
Solve[(Coefficient[f[k + 3], f[1]]*Coefficient[f[k + 1], f[3]] -

\QTP{Body Math}
$\qquad \ \ \ $ Coefficient[f[k + 3], f[3]]*Coefficient[f[k + 1], f[1]]) ==
0, E1]/. $\alpha \rightarrow $ 0.1

\QTP{Body Math}
$\qquad $(*gives eigenvalues of the odd states*)

It is obvious that, the program can easily be adapted for similar problems.
The method introduced here gives accurate results for bosonic Hamiltonian (%
\ref{eq:6}). The results are given in Table I. As shown in the Table I our
data confirm some previous results. Note that the results are obtained after 
$20$ iteration. 
\begin{table}[tbph]
\begin{tabular}{llll}
\hline\hline
$n$ \ \ \ \ \ \ \  & $E_{present}$ \ \ \quad \qquad & $E$ \cite{ciftci1}%
\qquad & $\qquad E$ \cite{must,koc8} \\ \hline\hline
$0$ & $1.065286$ & $1.065286$ & $\qquad 1.065286$ \\ 
$1$ & $3.306872$ & $3.306871$ & $\qquad 3.306872$ \\ 
$2$ & $5.747959$ & $5.747960$ & $\qquad 5.747959$ \\ 
$3$ & $8.352678$ & $8.352642$ & $\qquad 8.352678$ \\ 
$4$ & $11.09860$ & $11.09835$ & $\qquad 11.09860$ \\ 
$5$ & $13.96993$ & $13.96695$ & $\qquad 13.96993$ \\ \hline\hline
\end{tabular}%
\caption[Different values of $s$ for $A=B=1$ and the energy values for type-1%
]{The comparison of eigenvalues of anharmonic oscillator computed by the AIM 
\protect\cite{ciftci1}, direct numerical integration method \protect\cite%
{must,koc8} and by the present work, ATEM when $\protect\alpha =0.1$.}
\end{table}

In the following subsections, it is shown that this asymptotic approach
opens the way to the treatment of \ single boson quantum optical systems.

\subsubsection{A simple multiphoton interaction Hamiltonian}

Hamiltonian of the single mode coherent light with an optically bistable two
photon medium is given by \cite{gerry,fabio,wu}%
\begin{equation}
H=\omega a^{+}a+\kappa \left( a^{+2}-a^{2}\right) +\Omega a^{+2}a^{2}
\label{eq:13}
\end{equation}%
where $\omega $\ is frequency, and $\kappa $\ and $\Omega $\ are real
constants. Time development of the Hamiltonian (\ref{eq:13}) was studied by 
\cite{gerry}. Here we study the determination of the eigenstate of the
equation $H\left\vert n\right\rangle =E\left\vert n\right\rangle .$ The
action of the Hamiltonian on the state $\left\vert n\right\rangle $ can be
written as%
\begin{equation}
\left( \omega n+\Omega n(n-1)-E\right) \left\vert n\right\rangle +\kappa
\left( \sqrt{n(n-1)}\left\vert n-2\right\rangle -\sqrt{(n+1)(n+2)}\left\vert
n+2\right\rangle \right) =0.  \label{eq:14}
\end{equation}%
Our task is now to express $n^{th}$ state in terms of $\left\vert
0\right\rangle $ and $\left\vert 1\right\rangle $ states. 
\begin{eqnarray}
n &=&0;\ \left\vert 2\right\rangle =p_{0}\left\vert 0\right\rangle  \notag \\
n &=&1;\ \left\vert 3\right\rangle =p_{1}\left\vert 1\right\rangle  \notag \\
n &=&2;\ \left\vert 4\right\rangle =p_{2}\left\vert 0\right\rangle  \notag \\
&&\cdots  \label{eq:15} \\
n &=&m;\ \left\vert m+2\right\rangle =p_{m}\left\vert 0\right\rangle  \notag
\\
n &=&m+1;\ \left\vert m+3\right\rangle =p_{m+1}\left\vert 0\right\rangle . 
\notag
\end{eqnarray}

It is obvious that eigenvalues of (\ref{eq:13}) can be obtained for even/odd
eigenstates setting $p_{m}=0/$ $p_{m+1}=0.$ In this case we have used the
MATHEMATICA program code given in SECTION II. The results are given in Table
II. We have checked that the Hamiltonian (\ref{eq:13}) can exactly be solved
when $\Omega =0.$ In this case for $\kappa =\frac{\sqrt{3}}{2},$
eigenvalues, $E=2n+\frac{1}{2}$, and we have obtained the same result by
using the procedure given here.

\begin{table}[tbph]
\begin{tabular}{|l|l|l|l|l|}
\hline
$n$ & $\kappa =\frac{\sqrt{3}}{2};\Omega =0$ & $\kappa =0.1;\Omega =0.1$ & $%
\kappa =0.1;\Omega =0.5$ & $\kappa =0.5;\Omega =0.1$ \\ \hline
$0$ & $0.5$ & $0.00903368$ & $0.00665483$ & $0.19828087652$ \\ \hline
$1$ & $2.5$ & $1.02298633$ & $1.011994512$ & $1.52644677404$ \\ \hline
$2$ & $4.5$ & $2.23086041$ & $3.010484295$ & $2.9397418394$ \\ \hline
$3$ & $6.5$ & $3.63572596$ & $6.0102248594$ & $4.47732150123$ \\ \hline
$4$ & $8.5$ & $5.23894189$ & $10.010131290$ & $6.1677784947$ \\ \hline
$5$ & $10.5$ & $7.04117881$ & $15.010086374$ & $8.0292960950$ \\ \hline
\end{tabular}%
\caption[Different values of $s$ for $A=B=1$ and the energy values for type-1%
]{Eigenvalues $E$ of the Hamiltonian (\protect\ref{eq:13}), for $\protect%
\omega =1$ and various values of $\protect\kappa $ and $\Omega .$}
\end{table}
Consequently, we have shown that AIM can be applied to the determination of
the eigenstate of the single boson system.

\section{Eigenstate of multiboson Hamiltonians}

In this section we present application of the AIM in order to obtain
eigenvalues for a class of models describing two-mode multiphoton processes.
In addition to the annihilation operator $a,$ and creation operator $a^{+},$
we introduce the operators $b$ and $b^{+}$ in Hilbert space are given by 
\begin{subequations}
\begin{equation}
b^{+}=\frac{1}{\sqrt{2}}\left( -\frac{\partial }{\partial y}+y\right) ;\quad
b=\frac{1}{\sqrt{2}}\left( \frac{\partial }{\partial y}+y\right) .
\label{eq:3b}
\end{equation}
Two boson operator, $a$ and $b$, obey the usual commutation relations

\end{subequations}
\begin{equation}
\left[ a,b\right] =\left[ a,b^{+}\right] =\left[ b,a^{+}\right] =\left[
a^{+},b^{+}\right] =0,\quad \left[ a,a^{+}\right] =\left[ b,b^{+}\right] =1.
\label{eq:1}
\end{equation}%
Following a similar method which have been developed in the previous
section, we try to determine the eigenvalues for a general class of two-mode
multiphoton models. Hamiltonian of such system is given by 
\begin{equation}
H=r\omega a^{+}a+s\omega b^{+}b+\kappa (a^{+s}b^{r}+b^{+r}a^{s})
\label{eq:16}
\end{equation}%
where $r$ and $k$ are positive integers.

In this formalism when $r=s$ the Hamiltonian (\ref{eq:16}) satisfies the $%
SU(2)$ symmetry with the generators \cite{koc4,gursey}:%
\begin{equation}
J_{+}=a^{+}b,\quad J_{-}=b^{+}a,\quad J_{0}=\frac{1}{2}\left(
a^{+}a-b^{+}b\right) .  \label{eq:17}
\end{equation}%
These are the Schwinger representation of $su(2)$ algebra and they satisfy
the commutation relations

\begin{equation}
\left[ J_{+},J_{-}\right] =2J_{0}\quad \left[ J_{0},J_{\pm }\right] =\pm
J_{\pm }  \label{eq:18}
\end{equation}%
The fourth generator is the total boson number operator

\begin{equation}
N=(a^{+}a+b^{+}b)  \label{eq:19}
\end{equation}%
which commutes with the $su(2)$ generators. The Casimir operator of this
structure is given by

\begin{equation}
J=J_{-}J_{+}+J_{0}(J_{0}+1)=\frac{1}{4}N(N+2).  \label{eq:20}
\end{equation}%
If we denote the eigenvalues of the operator $J$ by

\begin{equation}
J=j(j+1)  \label{eq:21}
\end{equation}%
It is obvious that the irreducible representations of $su(2)$ can be
characterized by the total boson number $N=2j$. The application of the
realization (\ref{eq:17}) on a set of $2j+1$ states leads to the ($2j+1$%
)-dimensional unitary irreducible representation for each $j=0,1/2,1,....$
If the basis states are $\left\vert j,m\right\rangle $ ($m=j,j-1,...,-j$),
then the action of the operators on the basis states are given by:%
\begin{eqnarray}
J_{0}\left\vert j,m\right\rangle &=&m\left\vert j,m\right\rangle  \notag \\
J_{\pm }\left\vert j,m\right\rangle &=&\sqrt{(j\mp m)(j\pm m+1)}\left\vert
j,m\pm 1\right\rangle  \label{eq:22} \\
C\left\vert j,m\right\rangle &=&j(j+1)\left\vert j,m\right\rangle .  \notag
\end{eqnarray}%
An immediate practical consequence of these representation of $su(2)$
algebra is that the Hamiltonian (\ref{eq:16}) can easily be expressed as%
\begin{equation}
H=\omega sN+\kappa \left( J_{+}^{s}+J_{-}^{s}\right) .  \label{eq:23}
\end{equation}%
Eigenvalue equation $H\left\vert j,m\right\rangle =E\left\vert
j,m\right\rangle $ can be written as%
\begin{eqnarray}
&&\left( 2\omega sj-E\right) \left\vert j,m\right\rangle +  \notag \\
&&\kappa \sqrt{\frac{(-1)^{s}(m+s-j-1)!(m+s+j)!}{(m-j-1)!(m+j)!}}\left\vert
j,m+s\right\rangle +  \label{eq:24} \\
&&\kappa \sqrt{\frac{(-1)^{s}(-m+s-j-1)!(-m+s+j)!}{(-m-j-1)!(-m+j)!}}%
\left\vert j,m-s\right\rangle =0  \notag
\end{eqnarray}%
where $N=0,1,2,...$ .In this case the state $\left\vert j,m+s\right\rangle $
can be expressed as follows,%
\begin{eqnarray}
m=-j;\ &&\left\vert j,-j+s\right\rangle =p_{-j}\left\vert j,-j\right\rangle
+q_{-j}\left\vert j,-j-s\right\rangle  \notag \\
m=-j+1;\ &&\left\vert j,-j+s+1\right\rangle =p_{-j+1}\left\vert
j,-j+1\right\rangle +q_{-j+1}\left\vert j,-j-s+1\right\rangle  \notag \\
&&\cdots  \label{eq:25} \\
m=j-1;\ &&\left\vert j,j+s-1\right\rangle =p_{j-1}\left\vert
j,j-1\right\rangle +q_{j-1}\left\vert j,j-s-1\right\rangle  \notag \\
m=j;\ &&\left\vert j,j+s\right\rangle =p_{j-1}\left\vert j,j\right\rangle
+q_{j-1}\left\vert j,j-s\right\rangle ,  \notag
\end{eqnarray}%
boundary condition $\left\vert j,-j-s\right\rangle =0.$ The Hamiltonian (\ref%
{eq:23}) is exactly solvable when $s=1$ and a after some straightforward
treatment we can show that $E=2j+2(n-j)\kappa $. For the values $s=2$ and $%
j=3$, the the results are given in Table III. 
\begin{table}[tbph]
\begin{tabular}{|l|l|l|l|}
\hline
$m$ & $\kappa =\frac{1}{10};$ & $\kappa =\frac{1}{5};$ & $\kappa =\frac{1}{2}%
;$ \\ \hline
$0$ & $12$ & $12$ & $12$ \\ \hline
$\pm 1$ & $\frac{1}{5}\left( 57\pm 2\sqrt{6}\right) $ & $\frac{2}{5}\left(
27\pm 2\sqrt{6}\right) $ & $\left( 9\pm 2\sqrt{6}\right) $ \\ \hline
$\pm 2$ & $\frac{1}{5}\left( 60\pm 2\sqrt{15}\right) $ & $\frac{2}{5}\left(
30\pm 2\sqrt{15}\right) $ & $\left( 12\pm 2\sqrt{15}\right) $ \\ \hline
$\pm 3$ & $\frac{1}{5}\left( 63\pm 2\sqrt{6}\right) $ & $\frac{2}{5}\left(
33\pm 2\sqrt{6}\right) $ & $\left( 15\pm 2\sqrt{6}\right) $ \\ \hline
\end{tabular}%
\caption[Different values of $s$ for $A=B=1$ and the energy values for type-1%
]{Eigenvalues of the Hamiltonian (\protect\ref{eq:23}), for $\protect\omega %
=1,$ $s=2$ and $j=3.$}
\end{table}

\section{Conclusion}

The basic feature of our approach is to reformulate AIM for obtaining
eigenvalues of the bosonic Hamiltonians. Furthermore the technique given
here has been used to determine eigenvalues of anharmonic oscillator,
multiphoton interaction problem and a class of models describing two-mode
multiphoton processes. We have shown that AIM gives accurate results for
eigenvalue of bosonic Hamiltonians.

As a further work the method presented here can be developed in various
directions. Complete spectrum of the quasi-exactly solvable problems can be
obtained in the framework of the method presented here. Since most of the
quasi-exactly solvable problems can be expressed in terms of generators of $%
su(1,1)$ or $su(2)$ Lie algebra, the resulting recurrence relation can
easily be solved by using the procedure given in this paper. The suggested
approach can also be extended for solving boson-fermion systems. Before
ending this work a remark is in order. This extension leads to the
determination of eigenvalues of various Hamiltonians; Jahn-Teller
Hamiltonians \cite{koc5}, Rabi Hamiltonian \cite{koc6}, Hamiltonians of the
Bose-Einstein condensation problems.

\section{Acknowledgement}

The research was supported by the Scientific and Technological Research
Council of \ TURKEY (T\"{U}B\.{I}TAK).

\end{document}